\documentclass[fleqn,10pt]{wlscirep}

\usepackage{amsmath}
\usepackage{accents}
\usepackage[utf8]{inputenc}
\usepackage[T1]{fontenc}
\usepackage{authblk}
\usepackage{bm}
\usepackage{float}
\usepackage{breqn}

\newcommand{\be}{\begin{equation}}
\newcommand{\ee}{\end{equation}}
\newcommand{\bea}{\begin{eqnarray}}
\newcommand{\eea}{\end{eqnarray}}

\newcommand{\comment}[1]{}
\usepackage{xcolor,soul}
\setstcolor{red}

\title{Nanomagnetic Self-Organizing Logic Gates}

\author[1]{Pieter Gypens}
\author[1]{Jonathan Leliaert}
\author[3]{Massimiliano Di Ventra}
\author[1]{Bartel Van Waeyenberge}
\author[2,4]{Daniele Pinna}

\affil[1]{Dept. of Solid State Sciences, Ghent University, Belgium}
\affil[2]{Dept. of Physics, Johannes Gutenberg Universit\"at, Mainz, Germany}
\affil[3]{Department of Physics, University of California San Diego, La Jolla, USA}
\affil[4]{Peter Gr\"unberg Institute and Institute for Advanced Simulation, Forschungszentrum, J\"ulich, Germany}

\begin{abstract}
\end{abstract}

\begin{document}

\flushbottom
\maketitle
\thispagestyle{empty}

\textbf{The end of Moore's law for CMOS technology has prompted the search for low-power computing alternatives, resulting in several promising proposals  based on magnetic logic~\cite{allwood2005magnetic,LUO-20,DAT-90,sugahara2004spin,lee2007magneto, matsunaga2008fabrication,wang2005programmable,zhang2015magnetic}. One approach aims at tailoring arrays of nanomagnetic islands in which 
the magnetostatic interactions constrain the equilibrium orientation of the magnetization to embed logical functionalities~\cite{cowburn2000room, csaba2002nanocomputing, allwood2002submicrometer, imre2006majority}. 
Despite the realization of several proofs of concepts of such nanomagnetic logic~\cite{alam2010chip, breitkreutz2013experimental, eichwald2014majority}, it is still unclear what the advantages are compared to the widespread CMOS designs, due to their need for clocking~\cite{niemier2007clocking,bhowmik2014spin} and/or thermal annealing \cite{CAR-11, caravelli2020logical} for which fast convergence to the ground state is not guaranteed.
In fact, it seems increasingly evident that ``beyond CMOS'' technology will require a fundamental rethinking of our computing paradigm~\cite{conte2017rebooting}. In this respect, a type of \emph{terminal-agnostic} logic was suggested~\cite{DMM2}, where a given gate is able to ``self-organize'' into its correct logical states, regardless of whether the signal is applied to the traditional input terminals, or the output terminals. 
Here, we introduce \emph{nanomagnetic} self-organizing balanced logic gates, that employ stray-field coupled nanomagnetic islands to perform terminal-agnostic logic. 
We 
illustrate their capabilities by implementing reversible Boolean circuitry to solve a two-bit factorization problem via numerical modelling. In view of their design and mode of operation, we expect these systems to improve significantly over those suggested in Ref.~\cite{DMM2}, thus offering an alternative path to explore memcomputing,
whose usefulness has already been demonstrated by solving a variety of hard combinatorial optimization problems~\cite{DMMperspective}.
}

Magnetic logic is a promising alternative to the standard complementary metal–oxide semiconductor (CMOS) technology. To date, this paradigm has brought us domain wall logic~\cite{allwood2005magnetic,LUO-20},  spintronic field-effect transistors~\cite{DAT-90,sugahara2004spin}, magnetic tunnel junctions~\cite{lee2007magneto, matsunaga2008fabrication} reprogrammable magnetic random access memory cells~\cite{wang2005programmable}, and skyrmion logic devices~\cite{zhang2015magnetic}. 
Another promising approach is nanomagnetic logic ~\cite{cowburn2000room, csaba2002nanocomputing, allwood2002submicrometer, imre2006majority} (NML), where nanomagnetic islands interacting through their stray-field coupling are tailored to encode logic operations. However, their need for clocking~\cite{niemier2007clocking,bhowmik2014spin} and/or thermal annealing 
\cite{CAR-11, caravelli2020logical} impedes their use for \emph{reversible} logic operations. There, the entire logic circuit must relax to equilibrium as a whole, instead of through a cascade of  local gate relaxations. These difficulties are a well-known general property of quenching problems~\cite{gomes2008handbook} 
in complex effective energy landscapes~\cite{moore2011nature} where numerical solvers often employ cluster flipping operations to induce a long-range order in their energy minimization algorithms~\cite{swendsen1986replica, houdayer2001cluster }

As an alternative, a type of \emph{terminal-agnostic} logic was suggested in Ref.~\cite{DMM2}, where a given gate can dynamically ``self-organize'' into its logically correct states, irrespective of whether the signal is applied to the traditional input terminals, or the output terminals. These \emph{self-organizing logic gates} (SOLGs) form the elementary building blocks of \emph{digital memcomputing machines} ~\cite{DMM2}. These are machines that employ time non-locality (memory) to simultaneously process and store information. Their usefulness has already been demonstrated by solving a variety of hard combinatorial optimization problems, including, Boolean satisfiability (SAT), maximum satisfiability, integer linear programming, and even training of neural networks (see, e.g., Ref.~\cite{DMMperspective} for a brief review of these applications). The practical implementation of SOLGs suggested in Ref. ~\cite{DMM2} relies on resistive memories in addition to active devices, realizable with CMOS. Consequently, this design necessarily leads to an increased spatial footprint and sub-optimal energy consumption, two important aspects of any future computing paradigm.  

Here, we introduce a novel concept of SOLGs that instead employs stray-field coupled nanomagnetic islands to perform terminal-agnostic logic: \emph{nanomagnetic self-organizing logic} gates.  
Our strategy to build these gates relies on two main properties. First, we show that appropriately tailored stray-field interactions can enforce the logic proposition of the gate with equal population of all correct states, 
which is called \emph{balancedness}~\cite{LUS-99}. 
Second, we employ a 
\emph{dynamic error suppression},
to limit the time spent in excursions between logically  correct states, as a result of thermal fluctuations.

We will show that the combination of these two features is sufficient to implement a nanomagnetic SOLG, which can be used to construct reversible nanomagnetic SO-circuits. This will be demonstrated using a two-bit multiplier, which is capable of solving a simple factorization problem.

\vspace*{0.6cm}

Because mapping a whole computational problem to an arrangement of nanomagnetic islands whose energy landscape reflects the problem’s solution is far from trivial, we focus on the functionally complete NAND gate, from which \emph{any} Boolean circuit~\cite{kime2003logic} can be constructed in a bottom-up approach.

\begin{figure}[H]
	\centerline{\includegraphics[width=4in]{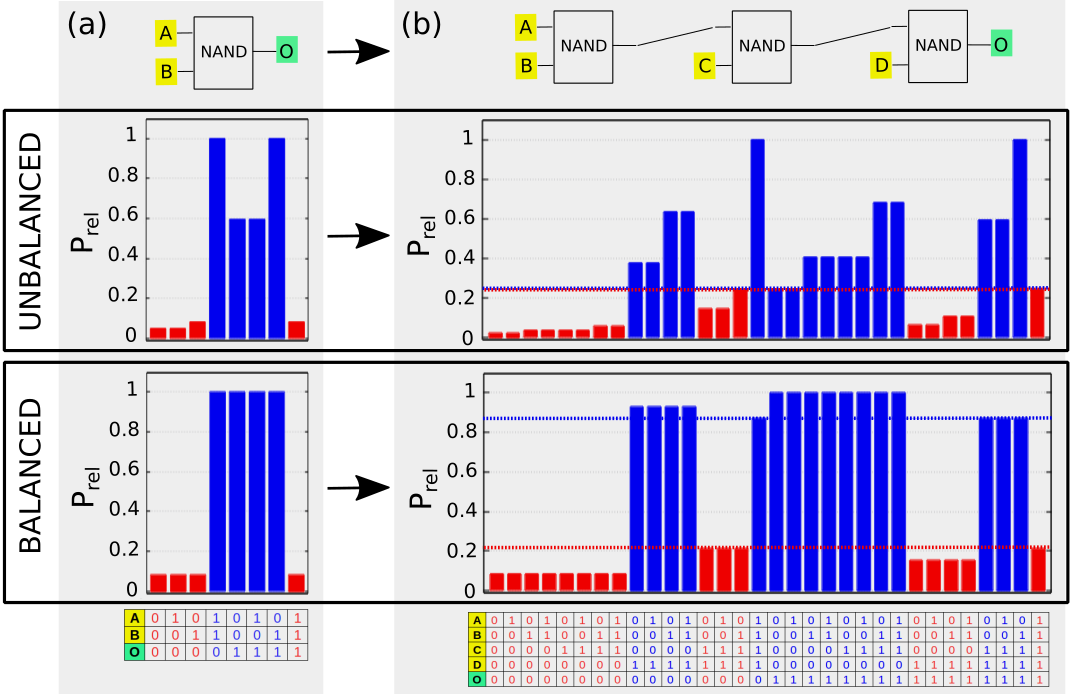}}
	\caption{{\footnotesize {\bf Effect of balancedness on logic circuit.} {\bf (a)} Boltzmann probability distribution for a unbalanced and a balanced NAND gate with a free output. $P_{{\mathrm rel}}$ is the probability relative to the most likely logical state. {\bf (b)} Boltzmann probability distribution for 3 NAND gates in series with a free output. The horizontal red line indicates the probability of the most likely logically incorrect state, while the horizontal blue line indicates the probability of the least likely logical state. If the circuit consists of unbalanced gates (upper panel), some logically correct states are indistinguishable from the logically incorrect states. In contrast, the same circuit built up out of balanced gates does not suffer from this problem (lower panel). 
	}}
	\label{fig:1_nand_design}
\end{figure}

For this type of logic to work, the NAND gate needs to be \emph{balanced}~\cite{LUS-99}, i.e.  
with equal probability for all logically correct states. 
To illustrate this, Figure \ref{fig:1_nand_design} shows how erroneous results arise when 3 unbalanced gates are integrated into a circuit. Even for a circuit as small as three gates, the least likely logically correct (horizontal blue line) and most likely logically incorrect state (horizontal red line) become indistinguishable, as opposed to the correct behaviour which is found when the circuit is assembled using balanced gates.

Recently, a balanced NAND gate consisting of 19 nanomagnetic islands with in-plane magnetization has been designed whose three lowest energy states present a four-fold degeneracy corresponding to the four NAND logic states~\cite{gypens2018balanced}. 

 This design, and any NML design that would be balanced following the definition given in Ref.~\cite{LUS-99}, i.e.\ that the energies of all \emph{relaxed} state are equal, has the drawback that it requires a slow and nontrivial annealing scheme to operate correctly since its equal population is only valid in the $T=0\,\mathrm{K}$ temperature limit.

\begin{figure}[H]
	\centerline{\includegraphics[width=2.5in]{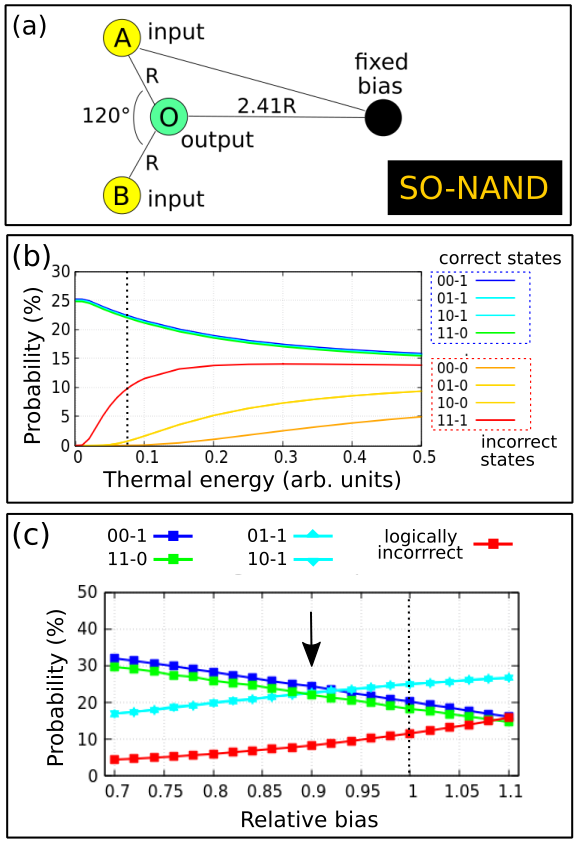}}
	\caption{{\footnotesize {\bf Balancedness of SO-NAND gate.} 
	{\bf (a)} Balanced SO-NAND gate consisting of four islands. The input islands (marked "A" and "B") are shown in yellow and the output (marked "O") in green. The black circle represent a bias island with fixed magnetization $m_z =-1$, which creates a bias field which is twice as large on the input islands as on the output island. 
	{\bf (b)} Boltzmann probabilities based on a two-state model (see Methods). Although the total probability of logically correct states decreases with increasing temperature (all states are equiprobable in the infinite temperature limit), their individual probabilities remain matched as the thermal energy is varied. 
	{\bf (c)} 
	The island macrospin dynamics are simulated at the thermal energy corresponding with the vertical dashed line in panel (b). The likelihood of each logical state is shown for varying relative magnetic moment of the fixed bias island (the two-state model corresponds to relative bias = 1). 
	Balancedness is recovered for a relative bias of 0.90, as indicated by the arrow. 
	}}
	\label{fig:2_balnand}
\end{figure}

For this gate design we can calculate the Boltzmann probabilities of the metastable state combinations corresponding
to the input/output islands magnetized (anti-)parallel to their uniaxial anisotropy axis. The Boltzmann probabilities as a function of thermal energy are shown in Figure \ref{fig:2_balnand}(b), demonstrating the balancedness of the gate design.

To investigate the gate's out-of-equilibrium behavior, this two-state model is extended to a dynamical macrospin model where the magnetizations can point in any direction and thermal fluctuations are taken into account. Although the additional degrees of freedom negatively impact the balancedness of the gate due to the enlarged phase space, it can be recovered by tuning the magnetic moment of the bias island, as shown in Figure \ref{fig:2_balnand}(c).
As shown in the supplemental materials, this tuning is robust with respect to variations in temperature and material parameters.

\vspace*{0.6cm}
Although the arrangement of nanomagnetic islands we have presented above gives rise to a balanced SO-NAND gate, the size of the circuits that can be constructed using this gate is limited by the fact that the probability of being in a logically incorrect state can be as high as 10\% for a single SO-NAND gate. To overcome this issue, we introduce a \emph{dynamic error suppression} (DES) scheme, which  suppresses the incorrect states while preserving the balancedness of the gate (Figure \ref{fig:2_NAND_overwiew}(a)). 

The DES scheme (see Supplementary Material for details and a possible physical implementation) is implemented in the model by periodically checking the logical correctness of the gate. If the gate is found to be in a logically incorrect state, an additional biasing field is applied on each input/output island for a small fraction (5\%) of the mean switching time of the biased gate, i.e.  $\tau_{\mathrm{on}} = 0.05 \tau_{\mathrm{switch}}$. These DES fields help to escape from an incorrect metastable state while still allowing the gate to explore different states. The rate at which this DES scheme is applied, ($\tau_{\mathrm{DES}}$), is tuned in order to find a trade off between the time the gate remains in a logically correct state and allowing the gate from accessing the full configuration space.

The most important aspect of the DES scheme is that it acts \emph{locally} at the level of each individual gate and the individual couplings, so that all gates in the circuit \emph{independently} suppress their respective logical errors. Whereas this may be reminiscent of the biasing used in p-bit logic inversion~\cite{borders2019integer}, for p-bit operation, the time-varying current through each sMTJ depends on the \emph{global} state of all the logical bits in the system. Without these globally-determined spin currents, however, no correct logical behavior would be observed as the sMTJ elements are not coupled in any other way. Stated differently, the p-bit logical functionality is \emph{entirely} embedded within the error suppression, and requires \emph{global} information. 
This is in sharp contrast to our approach, which only periodically suppresses logically incorrect states --which anyway are rare, as a consequence of the magnetostatic interactions-- using only \emph{local} information. For an individual gate, this is typically only applied for less then 1\% of the time.

Figure \ref{fig:2_NAND_overwiew}(c) shows how a circuit with the functionality of an SO-XNOR gate can be constructed by making use of SO-NAND, SO-NOR and a SO-NOT gates. The SO-NOT gate simply consists of two adjacent nanomagnetic islands, for which the minimum energy state corresponds to an anti-alignment of their magnetization. Interestingly, the symmetry of the SO-NAND gate dictates that when the magnetization of the fixed bias island is reversed [as shown in Fig. \ref{fig:2_NAND_overwiew} (b)], the same geometry behaves as a balanced SO-NOR gate, allowing us to use this same design to construct the SO-XNOR gate. 

The interactions between different gates are neglected, except for the islands which are part of the same coupling. These islands are magnetically coupled to align their magnetization. Aside from the DES biasing of individual gates, we introduce a modified DES scheme for such couplings, as detailed in the supplemental material.

\begin{figure}
	\centerline{\includegraphics[width=4.2in]{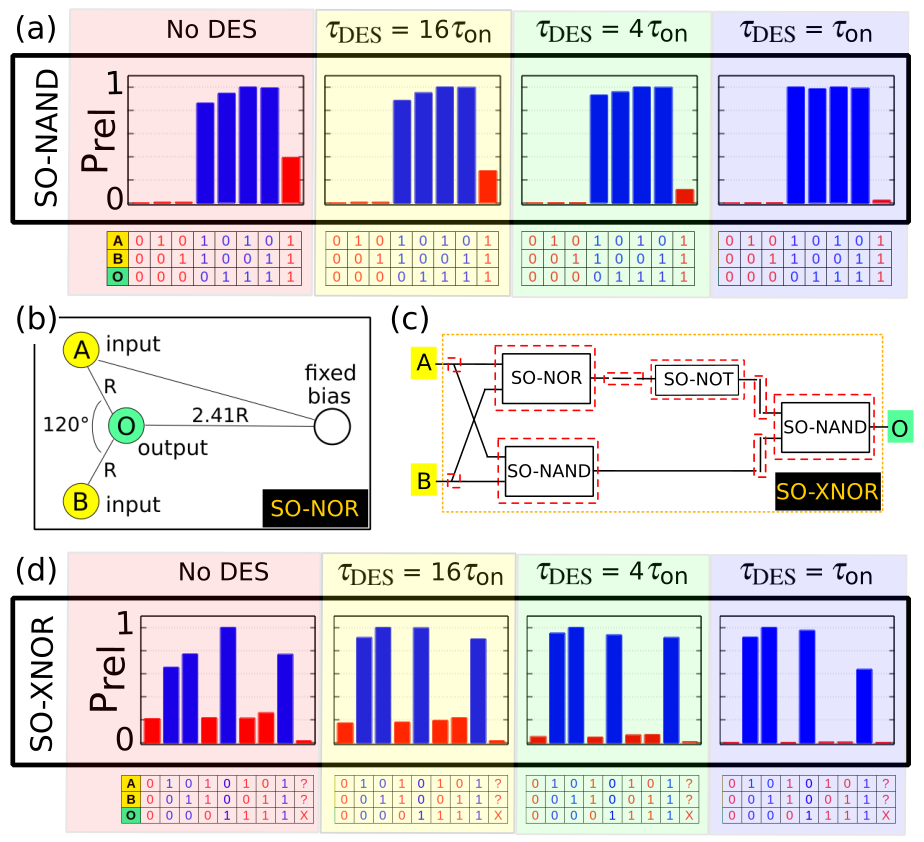}}
	\caption{{\footnotesize {\bf Performance of SO-NAND gate and SO-XNOR gate with dynamic error suppression (DES).} {\bf (a)} State probability distributions of a SO-NAND gate for different DES check times, $\tau_{\mathrm{DES}}$. $P_{{\mathrm rel}}$ is the probability relative to the most likely logical state. {\bf (b)} The SO-NOR gate is realized with the same geometry as the balanced SO-NAND gate, the only difference being the reversed magnetization of the fixed bias island. {\bf (c)} Schematic representation of a SO-XNOR gate. {\bf (d)} State probability distributions of a SO-XNOR gate for different $\tau_{\mathrm{DES}}$. The symbol ``?'' means that input islands A or input islands B are not logically consistent. All logically inconsistent states are aggregated into the X-state (X stands for 0 or 1).}}
	\label{fig:2_NAND_overwiew}
\end{figure}

\vspace*{0.6cm}

\begin{figure}
	\centerline{\includegraphics[width=\textwidth]{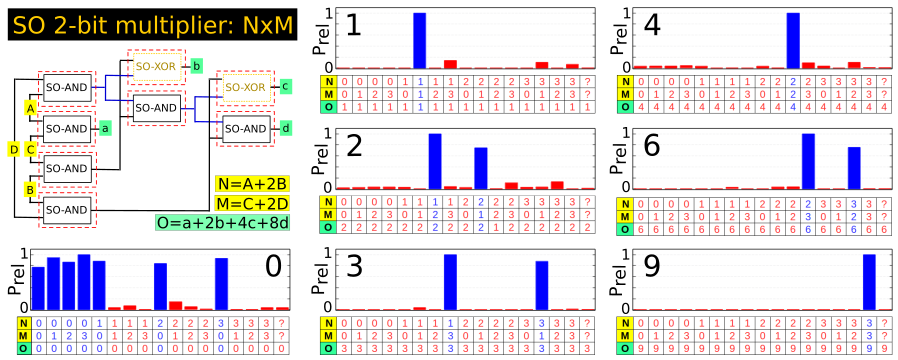}}
	\caption{{\footnotesize {\bf Operation of a self-organizing two-bit multiplier.} Schematic representation of a self-organizing two-bit multiplier and the state probability distributions for a fixed output, showing that for any fixed output we retrieve the correct factorization(s) with great probability. $P_{{\mathrm rel}}$ is the probability relative to the most likely logical state. All states of which one of the inputs is logically inconsistent are aggregated into the ``$?,?$''-state. 
	 The rate at which the DES scheme was applied, $\tau_{\mathrm{DES}}$, was set to $4\tau_{\mathrm{on}}$.
	}}
	\label{fig:4_2BM_overview}
\end{figure}

We will now demonstrate the use of our nanomagnetic SOLG in more complex logic circuits. In Figure \ref{fig:4_2BM_overview}, we present a self-organizing two-bit multiplier (SO-2BM) consisting of 46 nanomagnetic islands. As an example of a reverse computation, the four output islands of the SO-2BM, labeled from a to d, were fixed 
to represent specific numbers to be factorized (0, 1, 2, 3, 4, 6, or 9), and the system was allowed to explore its state space. The magnetization of the input islands, corresponding to the solution of our computation are labeled from A to D. As detailed in the supplemental materials, the balancedness of the gate has to be reconsidered when spins are fixed.
We demonstrate that our SO-2BM is capable of decomposing \emph{any} output number into its (prime)factors $-$ see Figure \ref{fig:4_2BM_overview}. Each possible factorization  has been found without clearly favoring nor penalizing any solution, which indicates that the balancedness remains largely conserved throughout the whole system. 

\vspace*{0.6cm}
In conclusion, we have shown that the use of balanced logic gates at nonzero temperature and \emph{local} dynamic error suppression can be leveraged to build \emph{nanomagnetic self-organizing logic gates}. These are \emph{terminal-agnostic gates} that can dynamically satisfy their logical proposition 
regardless of whether information is committed to the gate's inputs or outputs. They are the nanomagnetic equivalent of the building blocks required by digital memcomputing machines, which have already shown great promise in the solution of a variety of combinatorial optimization problems~\cite{DMMperspective}.

In particular, we have proposed a SO-NAND gate design that employs stray-field coupled perpendicularly-magnetized nanomagnetic islands.  
Because this gate is functionally complete, it allows to construct circuits in a bottom-up approach.

As an exemplary application, we have demonstrated number factorization by reversing the Boolean logic of a two-bit multiplier. 

It is worth stressing that this approach is fundamentally different from the one suggested in~\cite{sutton2017intrinsic,camsari2017stochastic,camsari2019p} where the magnetic states of Boltzmann machines have been designed to be constrained by spin-current biasing of isolated superparamagnetic tunnel junctions (sMTJ). Although this approach was notably demonstrated on a small, yet emblematic, prime factorization operation~\cite{borders2019integer}, it suffers a significant digital overhead because the individual current biases of all sMTJ elements in the circuit depend on the \emph{global} magnetic state, i.e.  the magnetic state of all elements. This means that the current biases need to be recalculated and adapted on the  timescale between two consecutive switches anywhere in the circuit\cite{borders2019integer}, which makes the architecture hard to scale. 
This contrasts our approach in which the \emph{local} DES scheme is applied on a timescale that only depends on the number of magnetic islands in an individual gate, allowing to build circuits of any size.

This means that nanomagnetic SOL gates may be employed in a wide variety of applications. Of particular interest are those related to cryptographically-important functions such as RSA~\cite{garg2016review}(Rivest–Shamir–Adleman) and ECDSA~\cite{johnson2001elliptic}(Elliptic Curve Digital Signature Algorithm), as well as many-to-one hashing functions~\cite{van1994parallel}, which are central to digital security~\cite{back2002hashcash} and blockchain protocols~\cite{nakamoto2019bitcoin}.

Finally, novel gate designs capable of inherently emulating more complex functionalities while requiring a smaller overall number of islands may exist. Finding such designs is, however, not a trivial task. Future work will attempt to use generative machine learning models to aid in the brute-force search of such design improvements. Regardless of these developments, the nanomagnetic self-organizing gates presented here represent a first important step towards the realization of unconventional computing architectures for the solution of a wide variety of problems of interest in academia and industry.

\clearpage
\section*{Methods}

{\bf Model Physics $-$} 
The magnetic free energy of the ensemble of $N$ nanomagnets is given by the combination of uniaxial magneto-crystalline anisotropy of each magnet plus the sum total of all pairwise stray-field interactions. Denoting by $\mathbf{M}$ the collection of all the rescaled magnetic moments $\mathbf{m}_i \equiv (m_{i;x},m_{i;y},m_{i;z})$, the total energy can be written:

\begin{equation}
\label{eq:energyD}
E(\mathbf{M}) = M_S^2 V \sum_i \left[-K \left(\mathbf{m}_i \cdot \mathbf{u}\right)^2 - \frac{\mu_0}{8\pi}\sum_{j \neq i} \left( \frac{3(\mathbf{m}_i \cdot \mathbf{r}_{ij})(\mathbf{m}_j \cdot \mathbf{r}_{ij})}{|\mathbf{r}_{ij}|^5}-\frac{\mathbf{m}_i \cdot \mathbf{m}_j}{|\mathbf{r}_{ij}|^3} \right) \right],
\end{equation}

where $M_S$, $V$, $K$, $\mathbf{u}$, $\mu_0$, and $\mathbf{r}_{ij}$ are the saturation magnetization, magnetic volume, magneto-crystalline anisotropy strength and direction, the magnetic permeability of free space and the pairwise distance vector $\mathbf{r}_{ij} \equiv \mathbf{r}_i-\mathbf{r}_j$ between islands $i$ and $j$, respectively.

We focus on magnetic islands that are magnetized out-of-plane at equilibrium, such that their easy-axis, $\mathbf{u}$, coincides with the $\mathbf{\hat{z}}$-axis. Our methodology would however also work for ensembles of in-plane magnetized islands. The anisotropic contribution to the energy in Eq. (\ref{eq:energyD}) is minimized locally for each macrospin of the ensemble when $m_{i,z}=\pm 1$. 

A number $N_{terminal}=N_{\mathrm{in}}+N_{\mathrm{out}} \leq N$ of spins in the ensemble will be denoted as {\it terminal} islands. 
Each such islands' perpendicular magnetization component is interpreted as a logical $0$ or $1$ depending on the sign of their $m_z$ component: $m_z < 0$ (bit $0$); $m_z > 0$ (bit $1$). Another group $N_{\mathrm{fix}}\leq N-N_{\mathrm{in}}-N_{\mathrm{out}}$ of islands will be denoted as fixed such that their magnetization imposes a static stray-field contribution on the rest of the ensemble. The remaining $N_{\mathrm{free}} = N-N_{\mathrm{in}}-N_{\mathrm{out}}-N_{\mathrm{fix}}$ islands (if any) will be free to fluctuate and dynamically evolve alongside the terminal islands. The evolution equations governing the magnetization dynamics of any non-fixed island in the ensemble will be given by the stochastic Landau-Lifshitz-Gilbert (sLLG) equation.
\begin{equation}
\label{eq:sLLG}
\dot{\mathbf{m}} = -\gamma\; \mathbf{m}\times\left[\frac{\nabla_{\mathbf{m}}E(\mathbf{M})}{\mu_0 M_S^2 V}+\mathbf{H}_{\mathrm{therm}}\right]+\alpha (\mathbf{m}\times\dot{\mathbf{m}})
\end{equation}

In this equation, $\dot{\mathbf{m}}$ denotes the time-derivative of the magnetization
, $\gamma$ the gyromagnetic ratio, $\alpha$ the unitless Gilbert damping constant.

Stochastic driving due to thermal noise is included in the $\mathbf{H}_{\mathrm{therm}}$ term. \\

{\bf Two-State model $-$} 
It is not computationally feasible to investigate every possible gate design by solving the full sLLG equation. Therefore, in our search for a suitable gate design (of which the presented SO-NAND gate is the result), we focused on solely evaluating the two-state behavior to find potential gate candidates $-$ see Fig.~\ref{fig:2_balnand}(b) where the Boltzmann probabilities of the eight possible states are plotted as a function of thermal energy. 

This analysis assumes that all free spins are in local equilibrium states $\mathbf{m}_i = (0,0,\pm 1)$ such that the magneto-crystalline anisotropy contributions in the first term of Eq. (\ref{eq:energyD}) are constant for any spin state and thus can be disregarded. Furthermore, as we consider islands located on the same lattice plane, the first term appearing in the stray-field contribution of Eq. (\ref{eq:energyD}) can also be discarded as it is identically zero. Computing the relative energy $\epsilon(\sigma_s)$ of each spin-flip state $\sigma_s$ can then be written as :

\begin{eqnarray}
\label{eq:dettime}
\epsilon(\sigma_s) = 
  \sum_{i,j=1,...,N} \left(\frac{\mu_0 M_S^2 V}{8\pi |\mathbf{r}_{i,j}|^3}\right) m_{i,z} m_{j,z},	
\end{eqnarray}
whence, the relative Boltzmann probability of any spin-flip state $\sigma_s$ at thermal equilibrium will be given by:

\begin{equation}
P[\sigma_s]= \frac{\exp^{-\beta \epsilon(\sigma_s)}}{\sum_s \exp^{-\beta \epsilon(\sigma_s)}}.
\end{equation}
where $\beta$ is the inverse thermal energy and the normalizing denominator runs over all $2^{(N-N_{\mathrm{fix}})}$ spin-flip states. 
The gate presented in Fig.~\ref{fig:2_balnand}(a) is optimal both with respect to the

difference in probability between logically correct states and the total probability of logically incorrect states.\\

\comment{
The probability of a given state $n$ can then be written as: 
\begin{equation}
\label{eq:prob} 
    P_n = \frac{{\rm exp}(E_n/k_{\rm B}T)}{\sum_{n=1}^{2^{N-N_{fix}}} {\rm exp}(E_n/k_{\rm B}T) }
\end{equation}
where $T$ is the temperature, $k_{\rm B}$ the Boltzmann constant, and the normalizing partition function is computed over all possible $2^{N-N_{fix}}$ spin-flip configurations. The energy of each state was calculated by
\begin{equation}
\label{eq:energy}
E = \frac{\mu_0 M_{{\rm s}}^2 V^2}{4\pi}\sum_{i \neq j} \frac{{m_z}_i {m_z}_j}{{R}_{ij}^3},
\end{equation}
with $M_{{\rm s}}$, $V$, $\mu_0$ and ${R}_{ij}$ being the saturation magnetization, magnetic volume, the magnetic permeability of free space and the pairwise distance between islands $i$ and $j$, respectively. 
}

{\bf Macrospin model $-$} The sLLG equation is solved using the macrospin simulation tool {\it Vinamax}\cite{LEL-15a}. 
In order to collect sufficient statistics about the state probability distribution within a reasonable computation time, we used simulation parameters that give rise to thermal switching dynamics on time scales approaching these of the magnetization dynamics: damping $\alpha$ = 0.4, temperature $T$ = 150 K, uniaxial anisotropy constant $K$ = 60 kJ/m$^3$, saturation magnetization $M_S$ = 1000 kA/m, magnetic volume $V$ = 180 nm$^3$, and $R$ = 9.0 nm. 

The statistics on the probability distribution for each spin state were collected by checking the logical state of the system each 10 ps for a period of 4 ms to 40 ms (depending on the size of the system). For the sake of simplicity, we used an initial state in which the magnetization of all free islands is set at $m_z = - 1$. However, the simulated time is sufficiently long to ensure that the results are independent of the initial state, and represent the probabilities in thermodynamic equilibrium.\\

\section*{Author contributions}
The project was conceived by JL and DP and supervised by JL, MD, BVW, and DP. The numerical experiments were performed and analysed by PG and DP. All authors discussed the results and wrote the manuscript.

\section*{Competing interests}
MD is the co-founder of MemComputing, Inc. 
(https://memcpu.com/) that is attempting to commercialize the
memcomputing technology. All other authors declare no competing interests.

\section*{Acknowledgements}
DP acknowledges the Helmholtz–RSF Joint Research Group "Exploring topological magnetization textures for artificial neural networks (TOPOMANN)", the Jülich Supercomputing Centre and RWTH Aachen  University for providing computational resources under project Nos. jiff40, and Dr. Karin Everschor-Sitte for partial funding support throughout the initial stages of this project. MD is supported by DARPA under grant No. HR00111990069. This  work  was  supported  by  the  Fonds  Wetenschappelijk  Onderzoek  (FWO-Vlaanderen) with a postdoctoral fellowship (JL). 

\bibliography{biblio}

\begin{thebibliography}{10}
\urlstyle{rm}
\expandafter\ifx\csname url\endcsname\relax
  \def\url#1{\texttt{#1}}\fi
\expandafter\ifx\csname urlprefix\endcsname\relax\def\urlprefix{URL }\fi
\expandafter\ifx\csname doiprefix\endcsname\relax\def\doiprefix{DOI: }\fi
\providecommand{\bibinfo}[2]{#2}
\providecommand{\eprint}[2][]{\url{#2}}

\bibitem{allwood2005magnetic}
\bibinfo{author}{Allwood, D.~A.} \emph{et~al.}
\newblock \bibinfo{journal}{\bibinfo{title}{Magnetic domain-wall logic}}.
\newblock {\emph{\JournalTitle{Science}}} \textbf{\bibinfo{volume}{309}},
  \bibinfo{pages}{1688--1692} (\bibinfo{year}{2005}).

\bibitem{LUO-20}
\bibinfo{author}{Luo, Z.} \emph{et~al.}
\newblock \bibinfo{journal}{\bibinfo{title}{Current-driven magnetic domain-wall
  logic}}.
\newblock {\emph{\JournalTitle{Nature}}} \textbf{\bibinfo{volume}{579}},
  \bibinfo{pages}{214--218} (\bibinfo{year}{2020}).

\bibitem{DAT-90}
\bibinfo{author}{Datta, S.} \& \bibinfo{author}{Das, B.}
\newblock \bibinfo{journal}{\bibinfo{title}{Electronic analog of the
  electro‐optic modulator}}.
\newblock {\emph{\JournalTitle{Applied Physics Letters}}}
  \textbf{\bibinfo{volume}{56}}, \bibinfo{pages}{665--667}
  (\bibinfo{year}{1990}).

\bibitem{sugahara2004spin}
\bibinfo{author}{Sugahara, S.} \& \bibinfo{author}{Tanaka, M.}
\newblock \bibinfo{journal}{\bibinfo{title}{A spin metal--oxide--semiconductor
  field-effect transistor using half-metallic-ferromagnet contacts for the
  source and drain}}.
\newblock {\emph{\JournalTitle{Applied Physics Letters}}}
  \textbf{\bibinfo{volume}{84}}, \bibinfo{pages}{2307--2309}
  (\bibinfo{year}{2004}).

\bibitem{lee2007magneto}
\bibinfo{author}{Lee, S.}, \bibinfo{author}{Choa, S.}, \bibinfo{author}{Lee,
  S.} \& \bibinfo{author}{Shin, H.}
\newblock \bibinfo{journal}{\bibinfo{title}{Magneto-logic device based on a
  single-layer magnetic tunnel junction}}.
\newblock {\emph{\JournalTitle{IEEE transactions on electron devices}}}
  \textbf{\bibinfo{volume}{54}}, \bibinfo{pages}{2040--2044}
  (\bibinfo{year}{2007}).

\bibitem{matsunaga2008fabrication}
\bibinfo{author}{Matsunaga, S.} \emph{et~al.}
\newblock \bibinfo{journal}{\bibinfo{title}{Fabrication of a nonvolatile full
  adder based on logic-in-memory architecture using magnetic tunnel
  junctions}}.
\newblock {\emph{\JournalTitle{Applied Physics Express}}}
  \textbf{\bibinfo{volume}{1}}, \bibinfo{pages}{091301} (\bibinfo{year}{2008}).

\bibitem{wang2005programmable}
\bibinfo{author}{Wang, J.}, \bibinfo{author}{Meng, H.} \&
  \bibinfo{author}{Wang, J.-P.}
\newblock \bibinfo{journal}{\bibinfo{title}{Programmable spintronics logic
  device based on a magnetic tunnel junction element}}.
\newblock {\emph{\JournalTitle{Journal of applied physics}}}
  \textbf{\bibinfo{volume}{97}}, \bibinfo{pages}{10D509}
  (\bibinfo{year}{2005}).

\bibitem{zhang2015magnetic}
\bibinfo{author}{Zhang, X.}, \bibinfo{author}{Ezawa, M.} \&
  \bibinfo{author}{Zhou, Y.}
\newblock \bibinfo{journal}{\bibinfo{title}{Magnetic skyrmion logic gates:
  conversion, duplication and merging of skyrmions}}.
\newblock {\emph{\JournalTitle{Scientific reports}}}
  \textbf{\bibinfo{volume}{5}}, \bibinfo{pages}{1--8} (\bibinfo{year}{2015}).

\bibitem{cowburn2000room}
\bibinfo{author}{Cowburn, R.} \& \bibinfo{author}{Welland, M.}
\newblock \bibinfo{journal}{\bibinfo{title}{Room temperature magnetic quantum
  cellular automata}}.
\newblock {\emph{\JournalTitle{Science}}} \textbf{\bibinfo{volume}{287}},
  \bibinfo{pages}{1466--1468} (\bibinfo{year}{2000}).

\bibitem{csaba2002nanocomputing}
\bibinfo{author}{Csaba, G.}, \bibinfo{author}{Imre, A.},
  \bibinfo{author}{Bernstein, G.~H.}, \bibinfo{author}{Porod, W.} \&
  \bibinfo{author}{Metlushko, V.}
\newblock \bibinfo{journal}{\bibinfo{title}{Nanocomputing by field-coupled
  nanomagnets}}.
\newblock {\emph{\JournalTitle{IEEE Transactions on Nanotechnology}}}
  \textbf{\bibinfo{volume}{1}}, \bibinfo{pages}{209--213}
  (\bibinfo{year}{2002}).

\bibitem{allwood2002submicrometer}
\bibinfo{author}{Allwood, D.} \emph{et~al.}
\newblock \bibinfo{journal}{\bibinfo{title}{Submicrometer ferromagnetic not
  gate and shift register}}.
\newblock {\emph{\JournalTitle{Science}}} \textbf{\bibinfo{volume}{296}},
  \bibinfo{pages}{2003--2006} (\bibinfo{year}{2002}).

\bibitem{imre2006majority}
\bibinfo{author}{Imre, A.} \emph{et~al.}
\newblock \bibinfo{journal}{\bibinfo{title}{Majority logic gate for magnetic
  quantum-dot cellular automata}}.
\newblock {\emph{\JournalTitle{Science}}} \textbf{\bibinfo{volume}{311}},
  \bibinfo{pages}{205--208} (\bibinfo{year}{2006}).

\bibitem{alam2010chip}
\bibinfo{author}{Alam, M.~T.} \emph{et~al.}
\newblock \bibinfo{journal}{\bibinfo{title}{On-chip clocking for nanomagnet
  logic devices}}.
\newblock {\emph{\JournalTitle{IEEE Transactions on Nanotechnology}}}
  \textbf{\bibinfo{volume}{9}}, \bibinfo{pages}{348--351}
  (\bibinfo{year}{2010}).

\bibitem{breitkreutz2013experimental}
\bibinfo{author}{Breitkreutz, S.} \emph{et~al.}
\newblock \bibinfo{journal}{\bibinfo{title}{Experimental demonstration of a
  1-bit full adder in perpendicular nanomagnetic logic}}.
\newblock {\emph{\JournalTitle{IEEE Transactions on Magnetics}}}
  \textbf{\bibinfo{volume}{49}}, \bibinfo{pages}{4464--4467}
  (\bibinfo{year}{2013}).

\bibitem{eichwald2014majority}
\bibinfo{author}{Eichwald, I.} \emph{et~al.}
\newblock \bibinfo{journal}{\bibinfo{title}{Majority logic gate for 3d magnetic
  computing}}.
\newblock {\emph{\JournalTitle{Nanotechnology}}} \textbf{\bibinfo{volume}{25}},
  \bibinfo{pages}{335202} (\bibinfo{year}{2014}).

\bibitem{niemier2007clocking}
\bibinfo{author}{Niemier, M.} \emph{et~al.}
\newblock \bibinfo{title}{Clocking structures and power analysis for
  nanomagnet-based logic devices}.
\newblock In \emph{\bibinfo{booktitle}{Proceedings of the 2007 international
  symposium on Low power electronics and design}}, \bibinfo{pages}{26--31}
  (\bibinfo{year}{2007}).

\bibitem{bhowmik2014spin}
\bibinfo{author}{Bhowmik, D.}, \bibinfo{author}{You, L.} \&
  \bibinfo{author}{Salahuddin, S.}
\newblock \bibinfo{journal}{\bibinfo{title}{Spin hall effect clocking of
  nanomagnetic logic without a magnetic field}}.
\newblock {\emph{\JournalTitle{Nature nanotechnology}}}
  \textbf{\bibinfo{volume}{9}}, \bibinfo{pages}{59} (\bibinfo{year}{2014}).

\bibitem{CAR-11}
\bibinfo{author}{Carlton, D.} \emph{et~al.}
\newblock \bibinfo{journal}{\bibinfo{title}{Computing in thermal equilibrium
  with dipole-coupled nanomagnets}}.
\newblock {\emph{\JournalTitle{IEEE Transactions on Nanotechnology}}}
  \textbf{\bibinfo{volume}{10}}, \bibinfo{pages}{1401--1404}
  (\bibinfo{year}{2011}).

\bibitem{caravelli2020logical}
\bibinfo{author}{Caravelli, F.} \& \bibinfo{author}{Nisoli, C.}
\newblock \bibinfo{journal}{\bibinfo{title}{Logical gates embedding in
  artificial spin ice}}.
\newblock {\emph{\JournalTitle{New Journal of Physics}}}
  \textbf{\bibinfo{volume}{22}}, \bibinfo{pages}{103052}
  (\bibinfo{year}{2020}).

\bibitem{conte2017rebooting}
\bibinfo{author}{Conte, T.~M.}, \bibinfo{author}{DeBenedictis, E.~P.},
  \bibinfo{author}{Gargini, P.~A.} \& \bibinfo{author}{Track, E.}
\newblock \bibinfo{journal}{\bibinfo{title}{Rebooting computing: The road
  ahead}}.
\newblock {\emph{\JournalTitle{Computer}}} \textbf{\bibinfo{volume}{50}},
  \bibinfo{pages}{20--29} (\bibinfo{year}{2017}).

\bibitem{DMM2}
\bibinfo{author}{Traversa, F.~L.} \& \bibinfo{author}{{Di Ventra}, M.}
\newblock \bibinfo{journal}{\bibinfo{title}{Polynomial-time solution of prime
  factorization and np-complete problems with digital memcomputing machines}}.
\newblock {\emph{\JournalTitle{Chaos: An Interdisciplinary Journal of Nonlinear
  Science}}} \textbf{\bibinfo{volume}{27}}, \bibinfo{pages}{023107}
  (\bibinfo{year}{2017}).

\bibitem{DMMperspective}
\bibinfo{author}{{Di Ventra}, M.} \& \bibinfo{author}{Traversa, F.~L.}
\newblock \bibinfo{journal}{\bibinfo{title}{Memcomputing: Leveraging memory and
  physics to compute efficiently}}.
\newblock {\emph{\JournalTitle{J. Appl. Phys.}}}
  \textbf{\bibinfo{volume}{123}}, \bibinfo{pages}{180901}
  (\bibinfo{year}{2018}).

\bibitem{gomes2008handbook}
\bibinfo{author}{Gomes, C.~P.} \emph{et~al.}
\newblock \bibinfo{journal}{\bibinfo{title}{Handbook of knowledge
  representation}}.
\newblock {\emph{\JournalTitle{Foundations of Artificial Intelligence}}}
  \textbf{\bibinfo{volume}{3}}, \bibinfo{pages}{89--134}
  (\bibinfo{year}{2008}).

\bibitem{moore2011nature}
\bibinfo{author}{Moore, C.} \& \bibinfo{author}{Mertens, S.}
\newblock \emph{\bibinfo{title}{The nature of computation}}
  (\bibinfo{publisher}{OUP Oxford}, \bibinfo{year}{2011}).

\bibitem{swendsen1986replica}
\bibinfo{author}{Swendsen, R.~H.} \& \bibinfo{author}{Wang, J.-S.}
\newblock \bibinfo{journal}{\bibinfo{title}{Replica monte carlo simulation of
  spin-glasses}}.
\newblock {\emph{\JournalTitle{Physical review letters}}}
  \textbf{\bibinfo{volume}{57}}, \bibinfo{pages}{2607} (\bibinfo{year}{1986}).

\bibitem{houdayer2001cluster}
\bibinfo{author}{Houdayer, J.}
\newblock \bibinfo{journal}{\bibinfo{title}{A cluster monte carlo algorithm for
  2-dimensional spin glasses}}.
\newblock {\emph{\JournalTitle{The European Physical Journal B-Condensed Matter
  and Complex Systems}}} \textbf{\bibinfo{volume}{22}},
  \bibinfo{pages}{479--484} (\bibinfo{year}{2001}).

\bibitem{LUS-99}
\bibinfo{author}{Lusth, J.~C.} \& \bibinfo{author}{Dixon, B.}
\newblock \bibinfo{journal}{\bibinfo{title}{A characterization of important
  algorithms for quantum-dot cellular automata}}.
\newblock {\emph{\JournalTitle{Information Sciences}}}
  \textbf{\bibinfo{volume}{113}}, \bibinfo{pages}{193--204}
  (\bibinfo{year}{1999}).

\bibitem{kime2003logic}
\bibinfo{author}{Kime, C.~R.} \& \bibinfo{author}{Mano, M.~M.}
\newblock \emph{\bibinfo{title}{Logic and computer design fundamentals}}
  (\bibinfo{publisher}{Prentice Hall}, \bibinfo{year}{2003}).

\bibitem{gypens2018balanced}
\bibinfo{author}{Gypens, P.}, \bibinfo{author}{Leliaert, J.} \&
  \bibinfo{author}{Van~Waeyenberge, B.}
\newblock \bibinfo{journal}{\bibinfo{title}{Balanced magnetic logic gates in a
  kagome spin ice}}.
\newblock {\emph{\JournalTitle{Physical Review Applied}}}
  \textbf{\bibinfo{volume}{9}}, \bibinfo{pages}{034004} (\bibinfo{year}{2018}).

\bibitem{borders2019integer}
\bibinfo{author}{Borders, W.~A.} \emph{et~al.}
\newblock \bibinfo{journal}{\bibinfo{title}{Integer factorization using
  stochastic magnetic tunnel junctions}}.
\newblock {\emph{\JournalTitle{Nature}}} \textbf{\bibinfo{volume}{573}},
  \bibinfo{pages}{390--393} (\bibinfo{year}{2019}).

\bibitem{sutton2017intrinsic}
\bibinfo{author}{Sutton, B.}, \bibinfo{author}{Camsari, K.~Y.},
  \bibinfo{author}{Behin-Aein, B.} \& \bibinfo{author}{Datta, S.}
\newblock \bibinfo{journal}{\bibinfo{title}{Intrinsic optimization using
  stochastic nanomagnets}}.
\newblock {\emph{\JournalTitle{Scientific reports}}}
  \textbf{\bibinfo{volume}{7}}, \bibinfo{pages}{1--9} (\bibinfo{year}{2017}).

\bibitem{camsari2017stochastic}
\bibinfo{author}{Camsari, K.~Y.}, \bibinfo{author}{Faria, R.},
  \bibinfo{author}{Sutton, B.~M.} \& \bibinfo{author}{Datta, S.}
\newblock \bibinfo{journal}{\bibinfo{title}{Stochastic p-bits for invertible
  logic}}.
\newblock {\emph{\JournalTitle{Physical Review X}}}
  \textbf{\bibinfo{volume}{7}}, \bibinfo{pages}{031014} (\bibinfo{year}{2017}).

\bibitem{camsari2019p}
\bibinfo{author}{Camsari, K.~Y.}, \bibinfo{author}{Sutton, B.~M.} \&
  \bibinfo{author}{Datta, S.}
\newblock \bibinfo{journal}{\bibinfo{title}{p-bits for probabilistic spin
  logic}}.
\newblock {\emph{\JournalTitle{Applied Physics Reviews}}}
  \textbf{\bibinfo{volume}{6}}, \bibinfo{pages}{011305} (\bibinfo{year}{2019}).

\bibitem{garg2016review}
\bibinfo{author}{Garg, S.} \& \bibinfo{author}{Rana, M.~K.}
\newblock \bibinfo{journal}{\bibinfo{title}{A review on rsa encryption
  algorithm}}.
\newblock {\emph{\JournalTitle{International Journal Of Engineering And
  Computer Science}}} \textbf{\bibinfo{volume}{5}} (\bibinfo{year}{2016}).

\bibitem{johnson2001elliptic}
\bibinfo{author}{Johnson, D.}, \bibinfo{author}{Menezes, A.} \&
  \bibinfo{author}{Vanstone, S.}
\newblock \bibinfo{journal}{\bibinfo{title}{The elliptic curve digital
  signature algorithm (ecdsa)}}.
\newblock {\emph{\JournalTitle{International journal of information security}}}
  \textbf{\bibinfo{volume}{1}}, \bibinfo{pages}{36--63} (\bibinfo{year}{2001}).

\bibitem{van1994parallel}
\bibinfo{author}{Van~Oorschot, P.~C.} \& \bibinfo{author}{Wiener, M.~J.}
\newblock \bibinfo{title}{Parallel collision search with application to hash
  functions and discrete logarithms}.
\newblock In \emph{\bibinfo{booktitle}{Proceedings of the 2nd ACM Conference on
  Computer and Communications Security}}, \bibinfo{pages}{210--218}
  (\bibinfo{year}{1994}).

\bibitem{back2002hashcash}
\bibinfo{author}{Back, A.} \emph{et~al.}
\newblock \bibinfo{title}{Hashcash-a denial of service counter-measure}
  (\bibinfo{year}{2002}).

\bibitem{nakamoto2019bitcoin}
\bibinfo{author}{Nakamoto, S.}
\newblock \bibinfo{title}{Bitcoin: A peer-to-peer electronic cash system}.
\newblock \bibinfo{type}{Tech. Rep.}, \bibinfo{institution}{Manubot}
  (\bibinfo{year}{2019}).

\bibitem{LEL-15a}
\bibinfo{author}{Leliaert, J.}, \bibinfo{author}{Vansteenkiste, A.},
  \bibinfo{author}{Coene, A.}, \bibinfo{author}{Dupr{\'e}, L.} \&
  \bibinfo{author}{Van~Waeyenberge, B.}
\newblock \bibinfo{journal}{\bibinfo{title}{Vinamax: a macrospin simulation
  tool for magnetic nanoparticles}}.
\newblock {\emph{\JournalTitle{Med. Biol. Eng. Comput.}}}
  \textbf{\bibinfo{volume}{53}}, \bibinfo{pages}{309--317}
  (\bibinfo{year}{2015}).

\end{thebibliography}

\end{document}